# Liquid-State Semiconductor Lasers Based on Type-(I+II) Colloidal Quantum Dots

Donghyo Hahm[1], Valerio Pinchetti[1], Clément Livache[1], Namyoung Ahn[1], Jungchul Noh[1], Xueyang Li[2], Jun Du[1,2], Kaifeng Wu[1,2], and Victor I. Klimov[1]*

[1]Nanotechnology and Advanced Spectroscopy Team, C-PCS, Chemistry Division, Los Alamos National Laboratory, Los Alamos, NM 87545, USA

[2]State Key Laboratory of Molecular Reaction Dynamics, Dalian Institute of Chemical Physics, Chinese Academy of Sciences, Dalian, Liaoning 116023, China

*klimov@lanl.gov

**Present-day liquid-state lasers are based on organic dyes[1]. Due to their large optical-gain bandwidth, they are often utilized in applications that require a tunable lasing wavelength. They are also applied in areas that necessitate the use of liquid optical gain media such as optofluidics and lab-on-a-chip diagnostics. To circumvent interference from nonemissive triplet states and instabilities under intense pumping, laser dyes usually employ free-floating jet streams. This complicates the system, increases its footprint, and makes it prone to malfunction due to accidental splashing. Here we demonstrate an alternative class of liquid-state lasers that overcome the shortcomings of traditional dye-based systems. These dye-like lasers employ solutions of semiconductor nanocrystals known as colloidal quantum dots (QDs)[2,3]. Previous efforts to realize such devices have been hampered by fast nonradiative Auger recombination of multi-carrier states needed for optical gain[4,5]. We overcome this challenge using type-(I+II) QDs that feature a trion-like optical-gain state with strongly suppressed Auger recombination. When combined with a Littrow optical cavity, 'static' (non-circulated) solutions of these QDs exhibit stable lasing tunable from 634 nm (red) to 594 nm (orange). Our findings demonstrate considerable promise of liquid-state lasers based on colloidal type-(I+II) QDs as a technologically viable alternative to traditional dye lasers. Importantly, the demonstrated QD lasers combine a large dye-like optical gain bandwidth with high operational stability as 'static' suspensions. Eliminating the need for fast circulation of a gain medium reduces the complexity (hence, the cost) of a laser system, allows for its miniaturization and simplifies its integration with various electronic and optical devices. An additional advantage of the QD gain media is unparallel diversity of their optical characteristics that can be readily manipulated by quantum-size effect, hetero-structuring, and compositional control. We anticipate that building on principles from the present study one will be able to realize liquid-state QD lasers with on-demand wavelengths spanning from infrared to ultraviolet spectral ranges.**

Chemically prepared semiconductor nanocrystals, or colloidal quantum dots (QDs), exhibit many properties that make them attractive for lasing applications[6-8]. These include a size-tunable emission wavelength, near-100% emission quantum yields, and a low optical gain threshold of around one electron-hole pair per dot on average. A natural form of colloidal QDs is a liquid suspension[3]. They are grown in a liquid medium and can be readily manipulated in solutions as big molecules. This makes them an ideal replacement for organic dyes in liquid-state lasers. However, initial efforts to realize lasing with solutions of ordinary QDs were not successful[6]. Only the use of dense solid-state QD films allowed for demonstrating amplified spontaneous emission (ASE)[5] and, eventually, lasing in devices supplemented by an optical cavity[9,10].

The reasons for the failure of earlier attempts to obtain lasing with QD solutions were uncovered in ref.[5], which established that the realization of laser action requires that a QD concentration ($n_{QD}$) in an optical gain medium must exceed a certain critical value ($n_{QD,cr}$). This was linked to peculiarities of the QD optical-gain mechanism which relies on stimulated emission from multi-carrier states. Indeed, because QD's band-edge states are at least two-fold spin-degenerate, the realization of optical gain requires the QDs to be populated with more than one electron-hole pair (one exciton) per-dot on average, that is, the QDs must contain an exciton and at least one additional carrier. The resulting complication is very fast optical gain decay arising from nonradiative Auger recombination during which an electron-hole recombination energy is transferred to a third carrier[4]. Typical time constants of this process ($\tau_A$) are tens to hundreds of picoseconds[4,11], which leads to very short gain lifetimes ($\tau_g$). For lasing to occur, the rate of stimulated emission ($r_{SE}$) must exceed that of optical gain decay ($r_g = 1/\tau_g \propto 1/\tau_A$), that is, $r_{SE} > r_g$. Since $r_{SE}$ scales directly with the QD concentration ($r_{SE} = \gamma n_{QD}$; $\gamma$ is a proportionality constant), stimulated emission overwhelms Auger decay only if $n_{QD}$ is greater than $n_{QD,cr} = (\gamma \tau_g)^{-1}$.

Very short Auger lifetimes lead to high values of $n_{QD,cr}$, which are difficult to obtain with QD solution samples due to solubility limits. Therefore, the original demonstration of ASE[5] as well as all validated reports of devices with confirmed lasing characteristics (including single-mode narrow-band emission, high temporal coherence, low-divergence output, and well-defined polarization) utilized solid-state QD films. While being highly desired, dye-like *technologically viable* liquid-state QD lasers have not been yet demonstrated.

Here, we realize these long-pursued devices using novel type-(I+II) QDs wherein optical gain is due to a multi-carrier state with charged-exciton-like characteristics. Among other gain-active species, a singly charged exciton has the fewest number of Auger decay pathways and, hence, exhibits the longest Auger time constant. To further extend the Auger lifetime, we grade the composition of the QD interior which helps produce a 'smooth', slowly varying carrier confinement potential. Such a potential is known to impede Auger decay due to suppression of the intra-band transition involving an energy-accepting carrier[12,13]. As a result, we achieve long optical gain lifetimes of more than 2.4 ns. This greatly reduces critical QD concentration $n_{QD,cr}$ and makes it accessible with solution-based samples. We incorporate a standard optical cell containing a 'static' (non-flowed) solution of the type-(I+II) QDs into a wavelength-tunable Littrow-type cavity with an optical grating as one of the reflectors. When this device is pumped with 2.33-eV, 5-ns pulses, it exhibits stable laser oscillations. The emitted light is characterized by a narrow spectral width down to 0.38 meV and a low divergence of 0.66 mrad or only 0.038º. By adjusting the blaze angle of the Littrow grating, we can tune the emission wavelength from 634 to 594 nm. No degradation of lasing power is observed for hours of operation without any agitation of the QD solution. These results demonstrate a great promise of solutions of type-(I+II) QDs as liquid-state optical gain media that combine a broad dye-like optical-gain bandwidth with advantageous features of inorganic low-dimensional semiconductors such as high operational stability and a wide range of accessible wavelengths readily selectable via established means for composition and size control.

**Optical gain due to hybrid biexcitons**

Most commonly, QD lasing occurs due to stimulated emission from neutral biexcitons (Fig. 1a). In this case, an optical gain lifetime is directly linked to the biexciton lifetime ($\tau_{XX}$). In standard (non-engineered) nanocrystals, it is dominated by nonradiative Auger recombination (that is, $\tau_{XX} \approx \tau_{A,XX}$), whose time constant follows well-established volume scaling ('$V$-scaling'), according to which $\tau_{A,XX} = \chi V_{QD}$, where $\chi$ is ~1 ps nm$^{-3}$ and $V_{QD}$ is the QD volume[6,11]. Based on $V$-scaling, for typical QD diameters from ~2 to ~6-nm $\tau_{A,XX}$ ranges from ~4 to ~100 ps. This leads to very short optical gain lifetimes, which greatly complicates the realization of lasing, especially in the case of continuous wave (*cw*) excitation or quasi-*cw* pumping with nanosecond or microsecond pulses[6].

The gain lifetime can be extended using charged excitons as optical gain species[14]. For example, due to the reduction in the number of Auger recombination pathways, the Auger lifetime of singly-charge excitons (trions) can be a factor of 4 (or more) longer than that of biexcitons[15]. A charged-exciton gain mechanism was first proposed in ref. [16], and recently it was employed to practically demonstrate 'zero-threshold' optical gain[14,17] and 'sub-single-exciton' lasing[18]. The reported studies utilized solid-state QD films that were charged electrochemically[16,17], photochemically[14,18], or chemically[19].

Here, we aim to realize a charged-exciton mechanism in solutions of QDs using specially engineered particles wherein biexcitons exhibit charged-exciton-like characteristics. The proposed QD design is depicted in Fig. 1b. Along with a primary confinement volume, which is common for an electron and a hole, it contains an additional spatially separated ('indirect') electron compartment whose interaction with the primary volume is controlled by a width and a height of an interfacial potential barrier. The electron energy in the indirect compartment should be slightly lower than that in the primary QD volume so a single electron-hole pair readily spatially separates to form an indirect exciton. In the case of a biexciton, the energetic driving force should still be enough to facilitate the transfer of one electron into the indirect compartment. However, it should not be too big, so the Coulombic effects (attraction to two holes in the primary volume and repulsion from an electron in the 'indirect' compartment) would prevent the transfer of the second electron. As a result, the biexciton will be stabilized in a 'hybrid' state composed of a direct and an indirect exciton.

The advantage of the hybrid biexciton versus an ordinary biexciton for lasing applications is an extended Auger lifetime. For the standard biexciton, Auger decay can be described by a superposition of four recombination pathways, two of which are associated with a positive trion and two with a negative trion (rates per pathway are $r_{A,X^+}$ and $r_{A,X^-}$, respectively)[15,20]. Hence, the biexciton Auger decay rate can be expressed as $r_{A,XX} = 2r_{A,X^+} + 2r_{A,X^-}$ (Fig. 1a and Supplementary Fig. 1a). For the hybrid biexciton, Auger dynamics are expected to be dominated by carriers residing in the primary QD volume. Such carriers form a positive-trion-like state whose decay occurs via a single Auger pathway. Hence, the decay rate is reduced to $r_{A,XX} = r_{A,X^+}$ (Fig. 1b and Supplementary Fig. 1b). Based on these considerations, we hypothesize that by implementing a hybrid-biexciton optical-gain scheme, we should be able to lengthen optical gain lifetimes and thereby realize lasing with solution forms of QDs.

**Type-(I+II) QDs**

To implement the hybrid-biexciton scheme (Fig. 1b), we have chosen a core/multi-shell design depicted in Fig. 2a (top). The proposed heterostructure comprises a CdSe core (primary confinement volume), a ZnSe shell (barrier layer), a CdS shell (indirect electron compartment), and a final ZnS layer which passivates the surface of the QD and enhances its stability. In the barrier layer, we grade the semiconductor composition using a $Cd_{1-x}Zn_xSe$ alloy wherein $x = 0$ at the core surface and $x$ approaches 1 at the shell periphery. This strategy allows us to alleviate a potential problem of crystalline defects that might develop due to larger directionally asymmetric lattice mismatch between CdSe and ZnSe. It further helps impede Auger recombination for both neutral and charge multi-carrier species[13] due to suppression of an intra-band transition involved in the dissipation of the electron-hole recombination energy[12]. For simplicity, further in this work, we will refer to the barrier layer as a 'ZnSe shell' or a 'ZnSe barrier'. Also, we do not account for the graded composition in the energy diagram in Fig. 2a.

The use of a wide-gap ZnSe shell allows us to create a large confinement potential for both an electron and a hole (Fig. 2a, bottom), as required by the scheme in Fig. 1b. CdS, used in the indirect-compartment layer, also possesses a bandgap which is wider than that of CdSe. Simultaneously, its conduction band edge is close to that of CdSe. Therefore, by varying the width of the CdS shell, it is possible to fine-tune the electron energy in the indirect compartment versus that in the core and thereby realize the desired regime when charge transfer is favored only for one electron of a biexciton state but not for two. This condition is critical for realizing a hybrid direct/indirect biexciton.

A synthetic approach to realize CdSe/ZnSe/CdS/ZnS QDs with controlled dimensions of all QD components is detailed in Methods. In Fig. 2b, we display a series of absorption and photoluminescence (PL) spectra of structures that occur at different stages of the growth of the targeted hetero-QDs. The PL measurements were conducted using low (sub-single-exciton) pump levels when the average excitonic number per QD, $\langle N \rangle$, is much less than 1. The transmission electron microscopy (TEM) images of the synthesized structures are shown in Supplementary Fig. 2.

First, we synthesize CdSe core particles with radius $r$ = 2.6 nm. They display a band-edge absorption peak at 2.11 eV and a spectrally symmetric PL band at 2.02 eV (top spectra in Fig. 2a). Then, we prepare a graded $Cd_{1-x}Zn_xSe$ shell. In this particular example, its thickness ($l$) is 1.7 nm. The deposition of the barrier layer leads to the development of a dual-peak structure in the band-edge absorption and asymmetric broadening of the PL band towards higher energies (second from the top spectra in Fig. 2a). These changes were observed previously and attributed to the increasing light-heavy-hole splitting due to asymmetric strain of the CdSe core[21]. The deposition of the CdS layer (thickness $h$ = 2.2 nm) leads to further broadening of the PL spectrum but now to lower energies (second from the bottom spectra in Fig. 2a). This signals the appearance of emission arising from a lower-energy indirect exciton formed by an electron in the CdS shell and a hole in the CdSe core. The double-hump PL structure becomes even more pronounced upon deposition of a final ZnS layer (thickness $d$ = 0.3 nm), which occurs due to increasing delocalization of the indirect exciton whose wave function extends into the outer shell (bottom spectra in Fig. 2a). As a result, the indirect exciton shifts to a lower energy which makes it more energetically distinct from the direct exciton. Based on a double-band deconvolution of the PL spectrum (Supplementary Fig. 3), the energies of the direct ($E_d$) and indirect ($E_i$) excitons are 2.02 eV and 1.95 eV, respectively.

The dual-exciton character of emission of the resulting hetero-QDs is also evident in single-dot spectra that exhibit a double-peak structure (Supplementary Fig. 4a). The two-state emission mechanism also manifests in double-exponential PL dynamics observed in both ensemble (Fig. 2c) and single-dot (Supplementary Fig. 4b) measurements which display distinct fast ($\tau_{f,X}$) and slow components ($\tau_{s,X}$) associated with direct ($X_d$) and indirect ($X_i$) excitons, respectively. In spectrally resolved PL time transients (Fig. 2d), the relative amplitudes of the fast and slow components correlate with energies of the $X_d$ and $X_i$ excitons. In particular, the fast signal is more pronounced in the spectral range of the higher-energy peak associated with the $X_d$ state.

To model the electronic states in our hetero-QDs, we solve the Schrödinger equation for electron and hole confinement potentials shown in Fig. 2a (bottom) taking into account electron-hole Coulomb interactions (Supplementary Note 1). In Supplementary Fig. 5, we display spatial distributions of electron and hole probabilities for the band-edge and the first excited excitonic states. In these calculations, we used QD dimensions that corresponded to those of the final QDs from Fig. 2b (bottom spectra). We observe that for the lowest-energy exciton, the electron and hole wavefunctions are primarily confined to, respectively, the CdS layer and the CdSe core (Supplementary Fig. 5a). This confirms our assessment that the band-edge transition manifested as a lower-energy hump in the PL spectrum corresponds to a spatially indirect exciton. The electron and hole wave functions of the next energetically distinct exciton state are co-localized in the core region (Supplementary Fig. 5b). This state, which can be classified as a 'direct exciton', is observed as a higher-energy hump in the PL spectrum. The calculated energies of the indirect and direct excitons ($E_i$ = 1.988 and $E_d$ = 2.037 eV, respectively) are in reasonable agreement with those inferred from the deconvolution of the measured PL spectrum (1.95 and 2.02 eV).

Usually, QDs are classified as type-I if their band-edge transition is associated with a direct exciton, and as type-II if the band-edge exciton is spatially indirect. In our structures, both type-I and type-II transitions are essential for understanding photophysical behaviors of band-edge excitons and biexcitons. Therefore, we dub our CdSe/ZnSe/CdS/ZnS heterostructures 'type-(I+II) QDs'.

To elucidate temporal characteristics of the $X_d$ and $X_i$ states, we conduct modeling of the measured PL dynamics and the overall (time-integrated) PL intensities using a scheme shown in the inset of Fig. 2c. Here, $\tau_d$ and $\tau_i$ are characteristic time constants of the transitions that couple, respectively, direct and indirect excitons to the ground state ('G' in the inset of Fig. 2c). Due to high emission quantum yields of our hetero-QDs (typically, 85% or higher), we assume that $\tau_d$ and $\tau_i$ are purely radiative time constants. We also account for bi-directional excitation transfer between the $X_d$ and $X_i$ states which we characterize by time constants $\tau_{d-i}$ ($X_d \rightarrow X_i$) and $\tau_{i-d}$ ($X_i \rightarrow X_d$). We further introduce probabilities $n_d$ and $n_i$ of an exciton to occupy either a direct or an indirect state ($n_d + n_i = 1$). Then, we solve two coupled rate equations to obtain temporal evolutions of $n_d$ and $n_i$ and use them to model PL dynamics and time-integrated PL intensities of the direct and indirect excitons (Supplementary Note 2).

This model allows us to quantitatively describe both the measured bi-exponential PL dynamics (Fig. 2c) and relative intensities of the $X_d$ and $X_i$ PL features. Based on the best correspondence between the measurements and the modeling, we obtain the following set of temporal characteristics of the coupled $X_d - X_i$ system: $\tau_d = 44$ ns, $\tau_i = 735$ ns, $\tau_{d-i} = 85$ ns, and $\tau_{i-d} = 8.8$ μs. As expected, the lifetime of the indirect exciton is considerably longer (factor of ~17) than that of the direct exciton. Further, a fairly slow $\tau_{d-i}$ time constant indicates that interstate coupling is not strong, which is a result of a large potential barrier separating the direct and indirect electron states. Further, the fact that $\tau_{d-i}$ is longer than $\tau_d$ is essential for explaining a discernable PL signal from the $X_d$ state, which is energetically higher than the $X_i$ state.

## Biexcitons in type-(I+II) QDs

Given the existence of two distinct band-edge excitons ($X_d$ and $X_i$), we consider the possibility of three different biexciton states – all-direct ($X_dX_d$), all-indirect ($X_iX_i$), and hybrid direct/indirect ($X_iX_d$) biexcitons. To determine energies of these states, we conduct quantum-mechanical effective-mass calculations detailed in Supplementary Note 1. Based on the modeling, the $X_iX_i$ and $X_iX_d$ biexcitons have comparable energies ($E_{id}$ = 4.024 and $E_{dd}$ = 4.020 eV, respectively), while the energy of the $X_dX_d$ state is considerably higher ($E_{dd}$ = 4.074 eV). This suggests that the biexcitonic response of our system should be dominated by two near-resonant $X_iX_i$ and $X_iX_d$ states. The fact that they are considerably closer in energy than the $X_d$ and $X_i$ excitons is a consequence of strong exciton-exciton repulsion characteristic of all-indirect biexcitons[22] which leads to a large high-energy shift of the $X_iX_i$ level versus that of two non-interacting $X_i$ excitons.

The energy of biexciton emission is defined by the difference of energies of the biexciton and the final single-exciton state. Since the $X_iX_d$ biexciton is expected to radiate primarily via the direct transition, its radiative decay should produce the $X_i$ exciton. The same final state is produced by radiative recombination of the $X_iX_i$ biexciton. Hence, the biexciton emission energies can be found from $h\nu_{ii} = E_{ii} - E_i$ and $h\nu_{id} = E_{id} - E_i$. These expressions yield $h\nu_{ii}$ = 2.032 eV and $h\nu_{id}$ = 2.036 eV. Both energies are close to the energy of the $X_d$ state (2.037 eV) because of which we expect that as a photoexcited system evolves from a single exciton to a biexciton state, a double-hump PL spectrum will be gradually replaced with single-band emission located near the $X_d$ PL feature.

Indeed, this trend is observed experimentally. In Fig. 3a, we show a series of pump-intensity-dependent PL spectra of an individual type-(I+II) QD. At low, sub-single-exciton pump levels, the single-dot PL line shape indicates the presence of two distinct transitions assigned earlier to the direct and indirect excitons. However, as the pump level is increased, the relative contribution of the lower-energy $X_i$ feature gets progressively decreased and the spectra become dominated by a single band whose energy is just slightly lower than the $X_d$ energy.

While the two biexcitonic components are not distinguishable in the PL spectra, they manifest as distinct components in PL time transients. To isolate biexciton decay from single-exciton dynamics, we apply two-photon correlation measurements conducted using the Hanbury Brown-Twiss (HBT) setup[23] (Methods). In this experiment, the emission from an individual QD is split between two channels equipped with single-photon detectors. Correlations between photon arrival events as a function of time difference between the two channels ($\tau$) yield information about second-order intensity correlation function $g^{(2)}(\tau)$. In the case of pulsed measurements, a nonzero $g^{(2)}$ signal at $\tau = 0$ implies that one pump pulse produced two photons, that is, the emitting state was a biexciton. Hence, the statistics of the delay of the first detected photon versus the pump pulse can be used to reconstruct biexciton dynamics.

Biexciton dynamics obtained by this approach (Fig. 3b) exhibits a two-component decay with 13.5 ns ($\tau_{f,XX}$) and 83 ns ($\tau_{s,XX}$) time, which we ascribe to the $X_iX_d$ and $X_iX_i$ states, respectively. Because the emission rate of the $X_iX_d$ state is much higher than that of the all-indirect $X_iX_i$ state, the hybrid direct/indirect biexciton is expected to be a primary optical gain state. If we neglect a weak contribution from the indirect transition, the number of radiative pathways for the $X_iX_d$ state is twice as large as for the single-exciton $X_d$ state, which yields radiative time constant $\tau_{id,rad} = 22$ ns[24]. The fast biexciton PL component is contributed by radiative and Auger decays of the $X_iX_d$ state and its population exchange with the $X_iX_i$ state. Hence, the overall recombination time of the $X_iX_d$ biexciton is at least 13.5 ns, based on which its emission quantum yield ($Q_{XX}$) is 61% (=13.5/22) or higher. Indeed, single-dot photon-correlation HBT measurements displayed in Fig.

3c yield $Q_{XX}$ of 71 %. The average of $Q_{XX}$ based on measurements of 26 QDs is even higher and reaches 76%. The HBT determination of $Q_{XX}$ employs the fact that the ratio of the $g^{(2)}$ amplitudes for the central ($\tau = 0$) and side ($\tau = T$) peaks is determined by the ratio of the biexciton and single-exciton PL quantum yields[25,26] ($T$ is the inter-pulse separation in a pump-pulse sequence).

The biexciton emission yield realized with our type-(I+II) QDs is almost twice as high as that observed previously for compositionally graded CdSe/CdZnSe QDs (cg-QDs) known for their excellent lasing performance[13,27]. This is a direct result of strong suppression of Auger decay caused by combined contributions of compositional grading and the reduction in the number of available Auger recombination pathways for the hybrid direct/indirect biexciton (Fig. 1).

## Optical gain in type-(I+II) QDs

Strong suppression of Auger decay in type-(I+II) QDs is anticipated to lead to long optical gain lifetimes which should facilitate lasing. To measure optical gain characteristics of the type-(I+II) QDs, we apply femtosecond transient absorption (TA) pump-probe measurements (Methods). TA spectra recorded as a function of pump level (Fig. 3d) reveal that optical gain (manifested as negative absorption) emerges at 1.96 eV, that is, near the peak of the $X_iX_d$ biexciton PL (Fig. 3a). As the pump intensity is ramped up, the width of the gain band increases and at $\langle N \rangle = 8.6$ it reaches ~150 meV, which is comparable to the bandwidth of organic laser dyes[1]. Based on the decay time of the negative absorption signal (Fig. 3e), optical gain lifetime is 2.4 ns, which is considerably longer than the gain lifespan observed previously for either regular or engineered QDs[6]. Interestingly, $\tau_g$ realized in our quantum confined structures (confinement energy is >200 meV) is comparable to the 3-ns gain lifetime observed recently for very large CdS nanocrystals wherein Auger recombination is absent due to a bulk-like character of electronic states[28].

## Liquid-state lasing with type-(I+II) QDs

The extended optical gain lifetime in type-(I+II) QDs translates into a reduced critical concentration required for lasing. In particular using $\tau_g$ = 2.4 ns in expression $n_{QD,cr} = (\gamma\tau_g)^{-1}$, we estimate $n_{QD,cr}$ of ~$10^{16}$ cm$^{-3}$ or ~18 µmol·L$^{-1}$ (Supplementary Note 3)[6,29]. Such concentrations are readily accessible with toluene solutions of the type-(I+II) QDs and, as shown below, using samples with $n_{QD}$ of more than ~10 µmol·L$^{-1}$ indeed allows us to attain laser action.

To realize lasing with a QD solution as an optical gain medium, we employ a Littrow-type wavelength-tunable cavity previously used in dye lasers[30] (Fig. 4a, Methods). It comprises a highly reflective planar mirror at one end and a reflection grating at the other end. The grating's blaze angle is set so that the first-order diffraction beam is reflected back into the cavity while the zero-order diffraction is used to outcouple light from the cavity. A 14 µmol·L$^{-1}$ QD solution is loaded into a 1-mm-thick quartz cuvette which is placed inside the cavity so as its wider 1-cm side is aligned with a cavity axis. The QDs are excited with 2.33-eV, 5-ns pulses from a neodymium-doped yttrium aluminum garnet (Nd:YAG) laser. The pump beam is focused onto a wider side of the cuvette into a 1-cm long horizontal stripe with a 200-µm width.

In Fig. 4b, we display the measurements of the output beam intensity as a function of per-pulse pump fluence ($J_p$) for the situation when the cavity is tuned to 1.984 eV. We detect no output signal up to $J_p$ of ~40 mJ cm$^{-2}$. When $J_p$ reaches 44 mJ cm$^{-2}$, an intense beam emerges from the cavity. It is highly directional and is detected as a bright spot with a 1.8-mm radius at a distance of 2.5 m from the cavity (Fig. 4c). Beam profile measurements yield divergence angle $\theta_d$ = 0.66 mrad (or only 0.038°) and beam waste $w_0$ = 0.28 mm (Fig. 4d).

As illustrated in the inset of Fig. 4b, the emitted light is characterized by a narrow linewidth of 2 meV (characterized in terms of a full width at half maximum, $\Gamma$). This value is validated by interferometric measurements of temporal coherence (Supplementary Fig. 6). In particular, using the measured coherence time ($\tau_c$ = 0.7 ps) in $\Gamma = h/(\pi\tau_c)$ ($h$ is the Planck constant), we obtain $\Gamma$ = 1.9 meV. By incorporating a beam expander into the cavity, we can further narrow the line to 0.38 meV or 1.2 Å (inset in Fig. 4e). This occurs due to the increasing size of the illuminated grating area which improves grating's spectral selectivity.

Polarization measurements (Supplementary Fig. 7) reveal that the emitted light is nearly perfectly linearly polarized in the direction perpendicular to grooves of the Littrow grating (polarization degree is ~95%). This is the result of the strong dependence of grating reflectively on polarization of incident light. In particular, the reflection coefficient for p-polarized light is >80%, versus 15% for s-polarized light.

A narrow linewidth of the emitted light, low beam divergence, and nearly perfect linear polarization unambiguously point toward the realization of the lasing regime. We can exclude that the observed effect is due to ASE as the ASE signal is observed as a broader, low-amplitude, spectrally distinct band whose threshold exceeds that of laser oscillations (Supplementary Fig. 8). To test for reproducibility of laser action, we have synthesized and studied three additional batches of the type-(I+II) QDs with parameters nominally identical to those of the dots in Fig. 4b-d. All additional samples show stable lasing for QD concentrations ranging from 14−30 µmol L$^{-1}$.

With a single QD sample, we can tune the lasing line from 1.96 to 2.03 eV (634 nm to 610 nm, respectively) by simply varying the grating blaze angle (Fig. 4e). We can extend the range of the covered spectral energies by exploiting tunability of the QD bandgap through particle size control.

To illustrate this capability, we prepare QDs with a CdSe core radius of 2.3 nm, which is smaller than that of the QDs discussed earlier. (The overall QD structure is CdSe ($r$ = 2.3 nm)/ZnSe ($l$ = 2.2 nm)/Cd$_{0.89}$Zn$_{0.11}$S ($h$ = 1.7 nm)/ZnS ($d$ = 0.3 nm)). As a result, we obtain a wider bandgap (2.12 eV versus 2.02 eV) which is expected to lead to bluer emission. The smaller-core-size QDs exhibit features typical of type-(I+II) structures including dual-band emission and two-component PL relaxation (Supplementary Fig. 9). These QDs also show strong lasing performance if incorporated in the Littrow cavity (Fig. 4e). Importantly, due to their wider bandgap, they allow us to access higher spectral energies and realize lasing tunable between 2.07 eV and 2.09 eV (599 nm and 594 nm, respectively). The total range of tunability achieved with two QD samples is from 1.96 eV to 2.09 eV, which spans 132 meV or 40 nm.

Another beneficial characteristic of the QDs is high operational stability achievable without sample agitation (by, *e.g.*, stirring or flowing). This is an important advantage over traditional laser dyes that require high-speed circulation (using, *e.g.*, a free-floating jet stream) for stable operation. In particular, our type-(I+II) QDs exhibit hours-long operational stability under static conditions while under the same conditions a standard laser dye (Rhodamine 6G) degrades within less than an hour (Supplementary Fig. 10).

To summarize, we demonstrate a new type of liquid-state lasers that employ colloidal type-(I+II) QDs. In these QDs, band-edge optical gain occurs due to hybrid direct/indirect biexcitons that display impeded trion-like Auger recombination. As a result, lasing becomes possible in liquid solutions of the QDs, for which it would be normally suppressed by very fast optical-gain relaxation caused by Auger decay. When incorporated into a Littrow-type optical cavity, the type-(I+II) QDs show narrow-line lasing tunable from 1.96 eV to 2.03 eV with a single QD sample. The use of wider bandgap QDs allows us to push lasing to higher spectral energies and reach 2.09 eV.

These findings demonstrate a considerable potential of type-(I+II) QDs as an immediate alternative to traditional laser dyes. Importantly, the QD lasers do not require circulation of a gain medium for stable operation. This should simplify system's design compared to existing dye lasers and reduce a device footprint. Additional benefits of QDs as liquid-state optical gain media stem from unmatched flexibility of their chemical properties (tunable by appropriate surface modifications) and optical characteristics (tunable via size, composition, and structure control). These features are expected to open the door to new applications of liquid-state lasers across diverse fields such as optofluidics, lab-on-chip diagnostics, high-contrast sensing and imaging, and many others.

**Methods**

**Chemicals.** Cadmium oxide (CdO, 99.5%, trace metals basis), zinc acetate (Zn(ac)$_2$, 99.99%, trace metals basis), oleic acid (OA, 90%, technical grade), 1-octadecene (ODE, 90%, technical grade), 1-octanethiol (OT, ≥ 98.5%), selenium (Se, ≥ 99.99%, trace metals basis) and sulfur (S, ≥ 99.0%) were purchased from Sigma-Aldrich. Tri-*n*-octylphosphine (TOP, 97%) was purchased from Strem Chemicals. All chemicals were used as received.

**Precursor preparation.** All chemical procedures were conducted in an inert atmosphere using the Schlenk line technique. Prior to the QD synthesis, stock solutions of 0.5 M zinc oleate (Zn(OA)$_2$), 0.5 M cadmium oleate (Cd(OA)$_2$), 2 M TOPSe, and 2 M TOPS were prepared. For the Zn(OA)$_2$ stock solution, a mixture of 100 mmol Zn(ac)$_2$ and a stoichiometric amount of OA were loaded into a flask and degassed under vacuum at 140 °C for 2 hours to achieve a clear solution. The flask was then purged with nitrogen, and the precursor concentrations were adjusted to 0.5 M using ODE. These precursor solutions were stored under an inert atmosphere at 100 °C for future use. To prepare the 0.5 M Cd(OA)$_2$ precursor solution, a combination of 20 mmol CdO, 20 ml OA, and 20 ml ODE was degassed in vacuum at 110 °C. The mixture was slowly heated to 300 °C to obtain a transparent solution. The reaction flask was subsequently cooled to 110 °C and vacuum-degassed again to eliminate any residual water. Stock solutions of 2 M TOPSe and TOPS were prepared by dissolving 100 mmol of Se and S in 50 ml of TOP at an elevated temperature. These stock solutions were then stored in a glove box for future use.

**Synthesis of type-(I+II) QDs.** Below, we describe a procedure used to prepare type-(I+II) QDs with the following structure: CdSe ($r$ = 2.6 nm)/ZnSe ($l$ = 1.7 nm)/CdS ($h$ = 2.2 nm)/ZnS ($d$ = 0.3 nm) QDs. We started the synthesis by loading 0.1 mmol of Cd(OA)$_2$ and 6 ml of ODE into a reaction flask. The mixture was degassed under vacuum at 110 ℃ and subsequently filled with nitrogen. After the mixture was heated to 310 ℃, 0.2 mmol of TOPSe was rapidly injected into the flask, immediately followed by a gradual dropwise addition of 1 ml of TOP. This process led to the formation of CdSe cores with $r = 1.5$ nm. A further increase of the CdSe core radius to 2.6 nm was achieved by the additional injection of Cd(OA)$_2$ (0.25 mmol) and TOPSe (0.25 mmol) precursors.

To grow a compositionally graded Cd$_{1-x}$Zn$_x$Se layer on top of the CdSe cores, a solution containing 0.4 mmol of Zn(OA)$_2$, 0.19 mmol of Cd(OA)$_2$ and 0.38 mmol of TOPSe was injected at an elevated temperature and allowed to react with the cores for 30 minutes. An additional dropwise injection of 1.31 mmol of Zn(OA)$_2$ and 1.52 mmol of TOPSe was carried out over 30 minutes, followed by an additional 30-minute reaction period to complete the Cd$_{1-x}$Zn$_x$Se barrier layer.

To grow a CdS layer, 4 mmol of OT was slowly added along with the stepwise injection of Cd(OA)$_2$. For the growth of an exterior ZnS shell, a mixture containing 2 mmol of OT and 4 mmol of Zn(OA)$_2$ was injected into the reaction flask. The temperature was then raised to 320 ℃ the reaction was allowed to proceed for one hour. Then, an additional 2 mmol amount of Zn(OA)$_2$ was added, and the reaction continued for 30 more minutes. After the completion of the reaction, the mixture was cooled to room temperature and the synthesized QDs were purified using precipitation/redispersion with acetone/toluene. Subsequently, the purified QDs were diluted with toluene for further use.

**Optical characterization.** Optical absorption and steady-state PL spectra of the synthesized QDs were measured using a Lambda 950 UV/VIS spectrometer (Perkin Elmer) and a Fluoromax+ spectrofluorometer (Horiba), respectively.

**Time-resolved photoluminescence measurements.** Power-dependent, time-resolved PL spectra were measured using a streak camera (Hamamatsu C10910). A continuously stirred QD sample in a 1-mm-thick quartz cuvette was excited by 343-nm (3.6 eV), 190-fs pulses of a tripled output of a regeneratively amplified ytterbium-doped potassium gadolinium tungstate (Yb:KGW) femtosecond laser (Pharos and HIRO, Light Conversion). The pulse repetition rate was 40 kHz. The PL was spectrally dispersed using a Czerny-Turner spectrograph (Acton 2300i) and sent into the streak camera unit to produce a two-dimensional (2D) PL intensity map with 'time' as a vertical axis and 'spectral energy' as a horizontal axis. Low-jitter, long-delay data were acquired by triggering the streak unit in a 'dump' mode, where the laser 40 kHz pulse picker signal was used to start the triggering sequence and the first available pulse in the 76 MHz oscillator train was used to generate a stable trigger signal. Using the Hamamatsu slow single-sweep unit, 2D (spectral energy-time) maps were acquired using different time ranges and incident pump fluences. In the case of the shortest time range used in the experiments (5 ns), the resolution of the streak camera was 12 ps, as inferred from the width of an instrument response function (IRF) obtain using a femtosecond laser pulse as the input signal.

**Transient absorption measurements.** Purified QDs were loaded into a 1-mm-thick quartz cuvette and stirred continuously. The TA measurements were performed using a pump–probe setup based on a regeneratively amplified Yb:KGW femtosecond laser (Pharos, Light Conversion) generating 190 fs pulses at 1030 nm with a 500 Hz repetition rate. Half of the laser fundamental output was used to seed a high-harmonic generator (HIRO, Light Conversion), producing second harmonic pulses (515 nm or 2.4 eV) used as the pump. The pump beam was modulated using an optical chopper synchronized such that to pick every other pulse from the pulse sequence. The pump pulses were focused into a 120–150 µm diameter spot onto the sample. The other half of the fundamental laser output at 1030 nm was fed into an optical delay line with an optical path varied from 0 ns to 4 ns. The delayed pulses were tightly focused onto a 5-mm-thick sapphire plate (EKSMA Optics) to generate a broadband white-light continuum. The generated white light was focused onto the sample into a 90 μm diameter spot in the middle of the pump spot. The actual pump and probe beam sizes were measured at the overlap using a beam profiler. The transmitted white light was detected using an Avantes AvaSpec-Fast ULS1350F-USB2 spectrometer.

Differential absorption spectra ($\Delta\alpha = \alpha_{\text{pump}} - \alpha_0$) were measured for each pump-probe delay; here $\alpha_{\text{pump}}$ and $\alpha_0$ are the absorption coefficients of the excited and unexcited sample, respectively. The correction for the probe *chirp* (spectro-temporal broadening) was performed following the procedure of ref. [31]. Excited-state absorption spectra were obtained from $\alpha = \alpha_0 + \Delta\alpha$. The realization of optical gain was indicated by the emergence of spectral regions wherein $\alpha$ was negative. Experimentally, this corresponded to the situation when sample excitation led to absorption bleaching ($\Delta\alpha < 0$) and the magnitude of the bleach signal was greater than that of linear absorption: $|\Delta\alpha| > \alpha_0$.

**Single-dot measurements**. For single-dot studies, QD samples were prepared as dilute sub-single-monolayer films (the QD areal density was ~0.01 per μm²) deposited onto cover slips via drop-casting. The samples were excited using 485 nm light generated by a PicoQuant LDH-D-C-485 laser diode. The laser operated at either 250 kHz or 500 kHz repetition rate (100 ps pulse duration) or in a *cw* mode. The excitation beam was focused onto the sample using an Olympus objective lens (50X, 0.70 NA). The PL signal was collected through the same objective lens. To measure single-dot PL spectra, QD emission was coupled into an imaging spectrometer (Acton Research, SpectraPro 500i) equipped with a charge coupled device (CCD) camera (Princeton Instruments,

PYLON). To measure PL dynamics, QD emission was spectrally filtered and detected by a pair of avalanche photodiodes (Micro Photon Devices, PDM Series) coupled to a start-stop time correlator (PicoQuant, HydraHarp 400). The temporal resolution of these measurements was 300 ps. A Hanbury Brown and Twiss configuration was employed to measure the second order intensity correlation function. All measurements were performed at room temperature under ambient conditions. The analysis of the experimental data was performed using custom-built software developed in Python.

**Liquid-state QD laser.** A solution sample of QDs in toluene with a concentration of about 14 µmol/L was loaded into a 1-cm wide, 1-mm thick quartz cuvette. The cuvette was placed inside a cavity composed of a high-reflectivity mirror and a reflection grating having 2400 groves per mm (Thorlabs). The wide side of the cuvette was aligned with the cavity axis. The grating angle was selected such that to send the 1$^{st}$-order diffraction beam back into the cavity (Littrow configuration). The 0$^{th}$-order diffraction was used to outcouple light from the cavity. The cavity resonance was tuned by rotating the grating assembly which allowed us to tune the output wavelength. Two intra-cavity lenses were used to collimate the beam and to ensure uniform illumination on the grating. The QD sample was excited from a wider side of the cuvette using 532-nm 5-ns second-harmonic pulses of a Nd:YAG laser (Amplitude Laser, Inc, Minilite II). The pump beam was shaped as a 1-cm long, 200-µm wide horizontal stripe using two cylindrical lenses.

The spectral analysis of the output beam of liquid-state QD lasers was performed using an imaging spectrometer (Acton Research, SpectraPro 500i) paired with a charge-coupled device (CCD) camera (Princeton Instruments, PYLON). The spectral resolution of this system was 120 µeV.

**Measurements of temporal coherence.** We employed a Michelson interferometer to measure temporal coherence of the QD laser output. In these measurements, the beam was split between two channels using a non-polarizing 50/50 beam splitter. Each channel was equipped with a flat mirror reflecting light back towards the beam splitter. One of the mirrors was mounted onto a single-axis linear delay stage (Aerotech ANT130L) which allowed us to control the time delay between the two paths. After passing the beam splitter, the two interfering beams were directed along the same path where they were collected using an Olympus PLN 10X objective used to couple light into a single-photon avalanche photodiode (MPD PDM series). To record the intensity of the interference pattern, we employed a time-tagged, time-resolved (TTTR) mode provided by a time correlated single photon counting (TCSPC) module (PicoQuant HydraHarp 400). The subsequent analysis of the TTTR data was performed using a Python code.

**Measurements of divergence.** The beam divergence was assessed by measuring the beam radius ($R_{beam}$) with a beam profiler (BladeCam-HR, DataRay). The light intensity profile was recorded as a function of distance from the cavity ($z$), and the beam halfwidth at the $1/e^2$ intensity level was used as a measure of a beam radius. The collected data were fitted to a hyperbolic function $R_{beam} = w_0(1 + z \tan(\theta))^{0.5}$, where $\theta$ is the divergence half-angle and $w_0$ is the beam waist radius.

**Data Availability**. The data that support the findings of this study are available from the authors on reasonable request.

1. Schäfer, F. P. *Dye lasers*. Vol. 1 (Springer Science & Business Media, 2013).
2. Brus, L. E. A simple model for the ionization potential, electron affinity, and aqueous redox potentials of small semiconductor crystallites. *J. Chem. Phys.* **79**, 5566-5571 (1983).
3. Murray, C. B., Norris, D. J. & Bawendi, M. G. Synthesis and characterization of nearly monodisperse CdE (E=S, Se, Te) semiconductor nanocrystallites. *J. Am. Chem. Soc.* **115**, 8706-8715 (1993).
4. Klimov, V. I., Mikhailovsky, A. A., McBranch, D. W., Leatherdale, C. A. & Bawendi, M. G. Quantization of multiparticle Auger rates in semiconductor quantum dots. *Science* **287**, 1011-1013 (2000).
5. Klimov, V. I. *et al.* Optical gain and stimulated emission in nanocrystal quantum dots. *Science* **290**, 314-317 (2000).
6. Park, Y.-S., Roh, J., Diroll, B. T., Schaller, R. D. & Klimov, V. I. Colloidal quantum dot lasers. *Nat. Rev. Mater.* **6**, 382-401 (2021).
7. Geiregat, P., Van Thourhout, D. & Hens, Z. A bright future for colloidal quantum dot lasers. *NPG Asia Mater.* **11**, 41 (2019).
8. Ahn, N., Livache, C., Pinchetti, V. & Klimov, V. I. Colloidal semiconductor nanocrystal lasers and laser diodes. *Chem. Rev.* **123**, 8251-8296 (2023).
9. Kazes, M., Lewis, D. Y., Evenstein, Y., Mokari, T. & Banin, U. Lasing from semiconductor quantum rods in a cylindrical microcavity. *Adv. Mater.* **14**, 317-321 (2002).
10. Eisler, H.-J. *et al.* Color-selective semiconductor nanocrystal laser. *Appl. Phys. Lett.* **80**, 4614-4616 (2002).
11. Robel, I., Gresback, R., Kortshagen, U., Schaller, R. D. & Klimov, V. I. Universal size-dependent trend in Auger recombination in direct-gap and indirect-gap semiconductor nanocrystals. *Phys. Rev. Lett.* **102**, 177404 (2009).
12. Cragg, G. E. & Efros, A. L. Suppression of Auger processes in confined structures. *Nano Lett.* **10**, 313-317 (2010).
13. Lim, J., Park, Y.-S. & Klimov, V. I. Optical gain in colloidal quantum dots achieved by direct-current charge injection. *Nat. Mater.* **17**, 42-49 (2018).
14. Wu, K., Park, Y.-S., Lim, J. & Klimov, V. I. Towards zero-threshold optical gain using charged semiconductor quantum dots. *Nat. Nanotech.* **12**, 1140-1147 (2017).
15. Wu, K., Lim, J. & Klimov, V. I. Superposition principle in Auger recombination of charged and neutral multicarrier states in semiconductor quantum dots. *ACS Nano* **11**, 8437-8447 (2017).
16. Wang, C., Wehrenberg, B. L., Woo, C. Y. & Guyot-Sionnest, P. Light emission and amplification in charged CdSe quantum dots. *J. Phys. Chem. B* **108**, 9027-9031 (2004).
17. Geuchies, J. J. *et al.* Quantitative electrochemical control over optical gain in quantum-dot solids. *ACS Nano* **15**, 377-386 (2021).
18. Kozlov, O. V. *et al.* Sub–single-exciton lasing using charged quantum dots coupled to a distributed feedback cavity. *Science* **365**, 672-675 (2019).
19. Whitworth, G. L., Dalmases, M., Taghipour, N. & Konstantatos, G. Solution-processed PbS quantum dot infrared laser with room-temperature tunable emission in the optical telecommunications window. *Nat. Photonics* **15**, 738-742 (2021).

20. Park, Y.-S., Bae, W. K., Pietryga, J. M. & Klimov, V. I. Auger recombination of biexcitons and negative and positive trions in individual quantum dots. *ACS Nano* **8**, 7288-7296 (2014).
21. Park, Y.-S., Lim, J. & Klimov, V. I. Asymmetrically strained quantum dots with non-fluctuating single-dot emission spectra and subthermal room-temperature linewidths. *Nat. Mater.* **18**, 249-255 (2019).
22. Klimov, V. I. *et al.* Single-exciton optical gain in semiconductor nanocrystals. *Nature* **447**, 441-446 (2007).
23. Brown, R. H. & Twiss, R. Q. Correlation between photons in two coherent beams of light. *Nature* **177**, 27-29 (1956).
24. Klimov, V. I. Multicarrier interactions in semiconductor nanocrystals in relation to the phenomena of Auger recombination and carrier multiplication. *Annu. Rev. Condens. Matter Phys.* **5**, 285-316 (2014).
25. Nair, G., Zhao, J. & Bawendi, M. G. Biexciton quantum yield of single semiconductor nanocrystals from photon statistics. *Nano Lett.* **11**, 1136-1140 (2011).
26. Park, Y. S. *et al.* Near-unity quantum yields of biexciton emission from CdSe/CdS nanocrystals measured using single-particle spectroscopy. *Phys. Rev. Lett.* **106** (2011).
27. Ahn, N. *et al.* Electrically driven amplified spontaneous emission from colloidal quantum dots. *Nature* **617**, 79-85 (2023).
28. Tanghe, I. *et al.* Optical gain and lasing from bulk cadmium sulfide nanocrystals through bandgap renormalization. *Nat. Nanotechnol.* **18**, 1423-1429 (2023).
29. Park, Y.-S., Bae, W. K., Baker, T., Lim, J. & Klimov, V. I. Effect of Auger recombination on lasing in heterostructured quantum dots with engineered core/shell interfaces. *Nano Lett.* **15**, 7319-7328 (2015).
30. Littman, M. G. & Metcalf, H. J. Spectrally narrow pulsed dye laser without beam expander. *Appl. Opt.* **17**, 2224-2227 (1978).
31. Livache, C. *et al.* High-efficiency photoemission from magnetically doped quantum dots driven by multi-step spin-exchange Auger ionization. *Nat. Photonics* **16**, 433-440 (2022).

**Acknowledgements.** This project was supported by the Laboratory Directed Research and Development (LDRD) program at Los Alamos National Laboratory under project 20230352ER. V.P. and N.A. acknowledge support by a LANL Director's Postdoctoral Fellowship. We thank Pamela R. Bowlan and Patrick J. Skrodzki for providing access to the Nd:YAG laser used as a pump source in the QD liquid-state experiments.

**Author Contributions.** V.I.K. initiated the study and together with D.H. conceived the idea of 'hybrid-biexciton' optical gain. He also analysed the data and coordinated project execution. D.H. and J.N. synthesized the type-(I+II) QDs and performed their structural analysis. C.L. and V.P. carried out TA and time-resolved PL measurements. D.H. performed single-dot spectroscopic measurements of the type-(I+II) QDs. He also conducted quantum-mechanical calculations of their electronic structures. X.L., J.D., and K.W. performed initial optical gain and amplified spontaneous emission studies of liquid QD suspensions using cg-QD samples. V.P., D.H., C.L., and N.A. developed the laser cavity design and performed liquid-state lasing experiments with type-(I+II) QDs. D.H. and V.I.K. wrote the manuscript with inputs from all co-authors.

**Competing interests.** The authors declare no competing interests.

**Additional Information.** Supplementary information is available in the online version of the paper. Reprints and permission information in available online at www.nature.com/reprints.

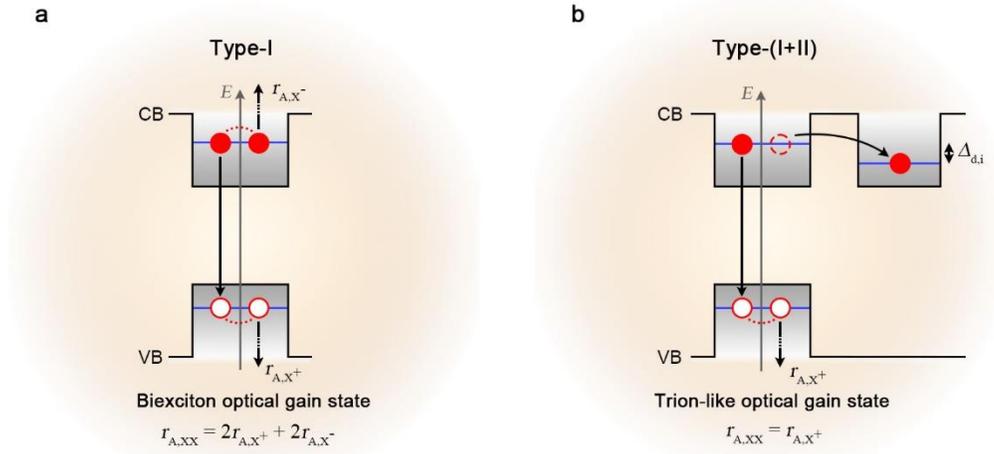

**Fig. 1 | Ordinary versus hybrid biexcitons. a,** In an ordinary QD with co-localized electrons and holes (type-I QD), optical gain is due to biexciton states. Their Auger recombination can be presented as a superposition of 2 negative- and 2 positive-trion pathways (rates $r_{A,X^-}$ and $r_{A,X^+}$ per pathway, respectively). CB and VB are conduction and valence bands, respectively. **b,** In the proposed scheme, optical gain is due to a 'hybrid biexciton' composed of a spatially direct and a spatially indirect exciton. Such biexcitons can be realized in so-called type-(I+II) QDs. These structures contain an additional, indirect compartment in the conduction band which captures an electron from a primary QD volume by providing a lower-energy state. The energy difference ($\Delta_{d,i}$), which drives electron transfer, however, must be fairly small to prevent transfer of the second electron that should remain in the primary volume. This would lead to the creation of a hybrid direct/indirect biexciton. Auger decay of such a biexciton occurs via a single positive-trion pathway. As a result, it is considerably slower than that of a regular biexciton.

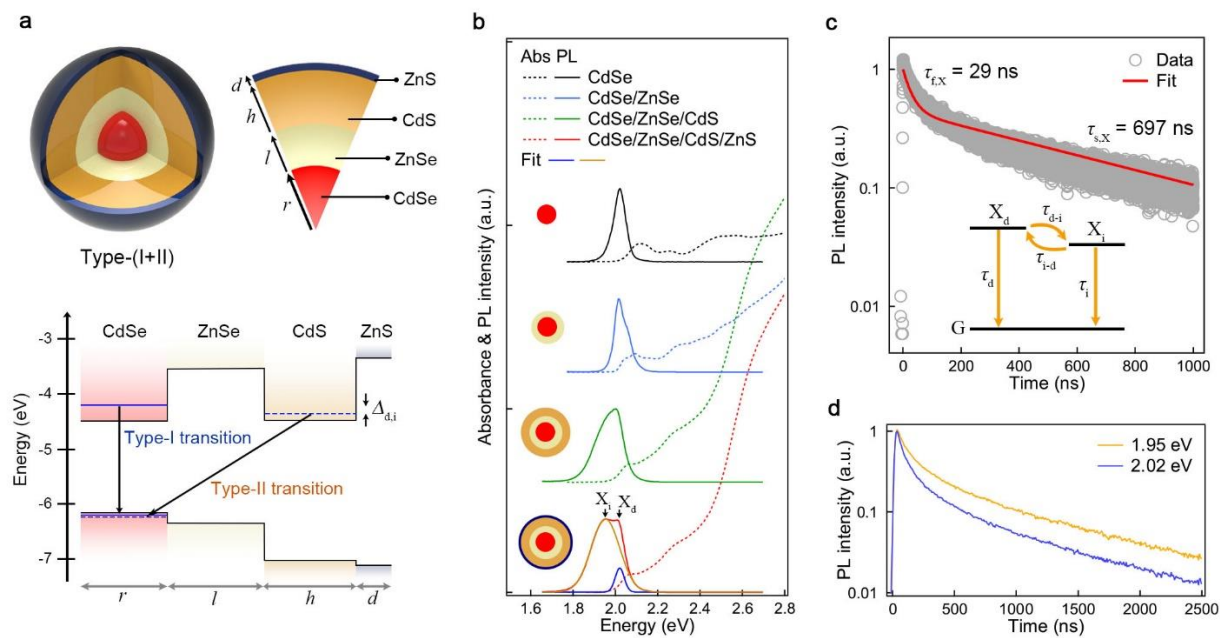

**Fig. 2 | Structural and optical characteristics of type-(I+II) QDs. a**, Top: A schematic depiction of a type-(I+II) CdSe/ZnSe/CdS/ZnS QD; $r$, $l$, $h$, and $d$ represent, respectively, the radius of the CdSe core, the thickness of the ZnSe barrier, the CdS interlayer thickness, and the thickness of the outer ZnS shell. Bottom: radial profiles of electron and hole confinement potentials in a type-(I+II) QD. It features spatially direct (type-I) and spatially indirect (type-II) transitions ($X_d$ and $X_i$ excitons, respectively). **b**, A series of absorption and PL spectra (dashed and solid lines, respectively) recorded at different stages of the synthesis of the type-(I+II) CdSe ($r$ = 2.6 nm)/ZnSe ($l$ = 1.7 nm)/CdS ($h$ = 2.2 nm)/ZnS ($d$ = 0.3 nm) QDs. Top-to-bottom: CdSe cores, CdSe/ZnSe core/barrier QDs, CdSe/ZnSe/CdS core/barrier/interlayer QDs, and CdSe/ZnSe/CdS core/barrier/interlayer/shell QDs. The PL spectrum of the final type-(I+II) QDs (red) can be presented as a sum of two Gaussians profiles centered at 2.02 eV and 1.95 eV. These bands correspond to the direct (blue) and indirect (gold) transitions, respectively. **c**, The measured spectrally integrated PL dynamics of the type-(I+II) QDs (grey circles) along with modeling (red line) conducted using a three-level scheme depicted in the inset. The best correspondence between the modeling and the experiment is obtained for the following parameters: $\tau_d$ = 44 ns, $\tau_i$ = 735 ns, $\tau_{d-i}$ = 85 ns, and $\tau_{i-d}$ = 8.8 μs. **d**, Spectrally resolved PL dynamics of the type-(I+II) QDs measured at 1.95 eV (yellow) and 2.02 eV (blue). The faster component is more pronounced in the trace recorded at 2.02 eV, which is near the center of the direct-exciton band.

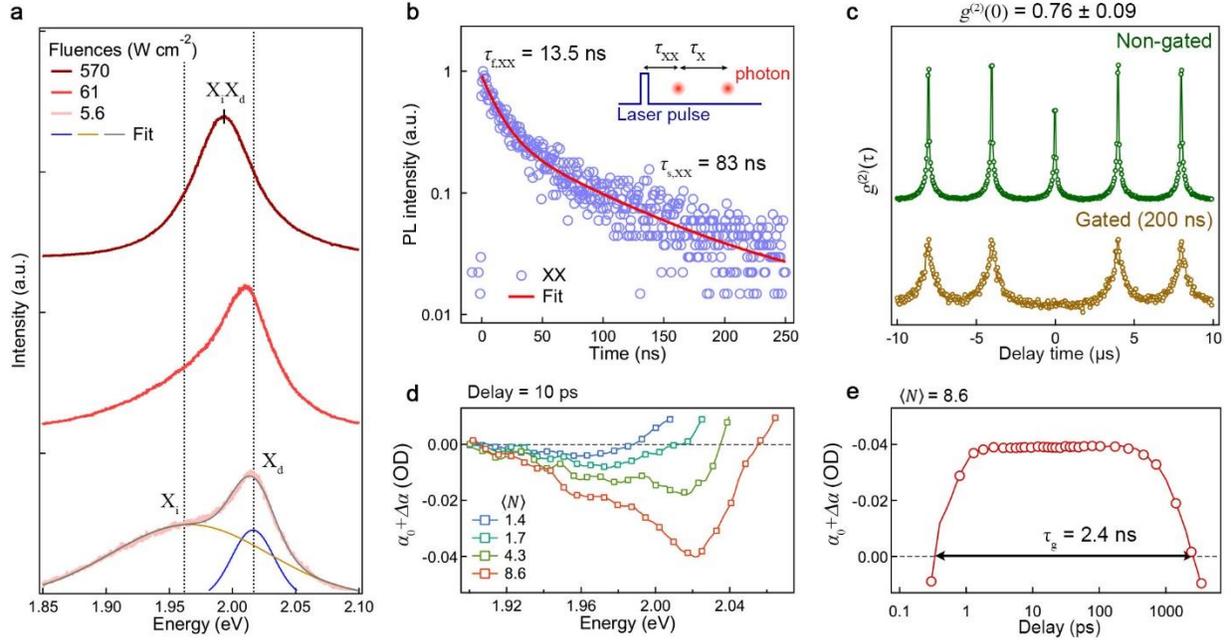

**Fig. 3 | Biexcitons in type-(I+II) QDs. a**, Pump-intensity-dependent PL spectra of a single type-(I+II) QD recorded using continuous wave (*cw*) 2.54-eV excitation. The lowest-pump-intensity spectrum is deconvolved into two bands that correspond to the direct ($X_d$) and indirect ($X_i$) excitons. This and other panels of this figure are for the same type-(I+II) QD sample as in Fig. 2. **b**, Single-dot measurements of biexciton PL decay (blue circles). These data were obtained using the Hanbury Brown-Twiss experiment, which allows one to identify excitation cycles with two emitted photons and to infer a biexciton lifetime from the delay of the first photon versus the pump pulse (inset). The red line is a two-exponential fit which yields the 13.5 ns and 83 ns time constants for, respectively, fast ($\tau_{f,XX}$) and slow ($\tau_{s,XX}$) relaxation components. **c**, Single-dot second-order intensity correlation ($g^{(2)}$) measurements without photon discrimination (top, green line) and with photon discrimination performed by selecting ('gating') photons with longer than 200-ns time delay versus the pump pulse (bottom, gold line). The 'non-gated' $g^{(2)}(0)$ averaged over 26 dots is 0.76 (inferred from the measurements without photon discrimination). This value yields the ratio of the biexciton and single-exciton emission quantum yields. The fact that the 'gated' $g^{(2)}(0)$ is close to zero confirms that the measurements were done for a single QD but not, *e.g.*, a cluster containing multiple QDs. **d**, Transient absorption spectra of type-(I+II) QDs presented as excited state absorption coefficient ($\alpha$) for the 10-ps pump-probe delay and for excitation levels $\langle N \rangle = 1.4$ to 8.6 (100-fs, 2.4-eV pump pulses). Optical gain corresponds to $\alpha < 0$. **e**, The excited state absorption dynamics yields optical gain lifetime $\tau_g = 2.4$ ns ($\langle N \rangle = 8.6$).

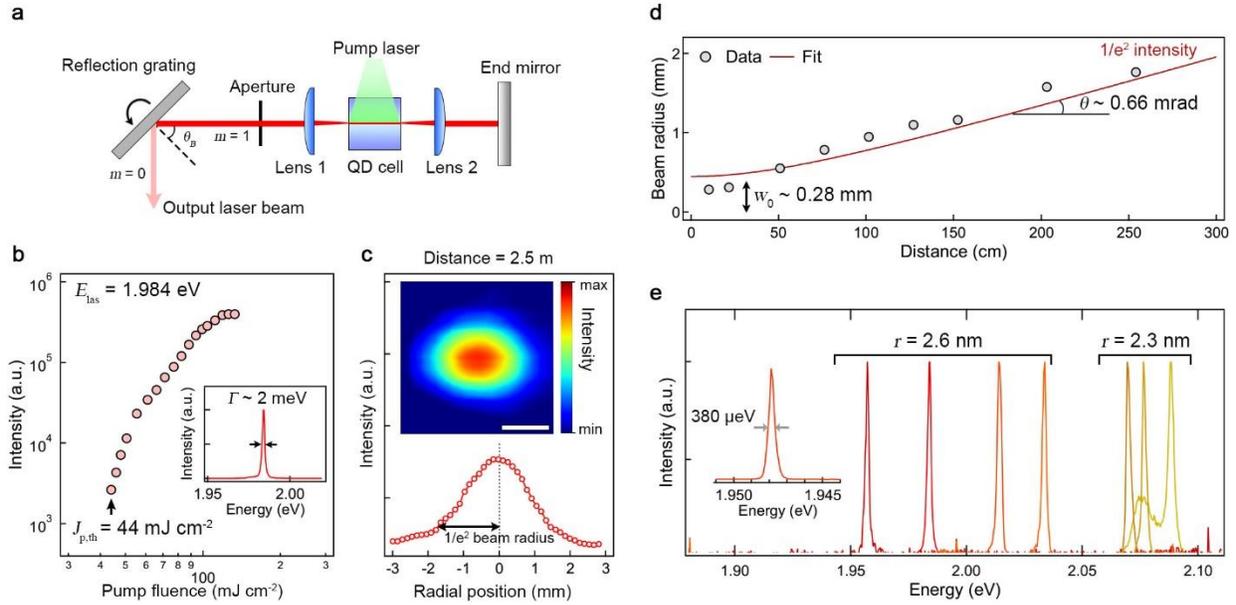

**Fig. 4 | Liquid-state lasing using type-(I+II) QDs. a**, A schematic diagram of a liquid-state laser with a solution of type-(I+II) QDs (14 μmol·L$^{-1}$) in an optical cell as an optical gain medium. This laser employs a wavelength-tunable Littrow cavity which comprises a planar mirror at one end and a reflection grating at the other end. The grating is positioned such that its first-order diffraction occurs in the direction of the incident beam (Littrow configuration). The zero-order diffraction is used to outcouple light from the cavity. Intracavity elements (two lenses and an aperture) help collimate the beam and improve its cross-section profile. **b**, The intensity of the output beam as a function of pump fluence for type-(I+II) QDs with $r = 2.6$ nm (same sample as in Figs. 2 and 3). The QD sample is pumped by 2.33-eV, 5-ns, 10-Hz pulses of the second harmonic of a Nd:YAG laser. The Littrow cavity is tuned to 1.984 eV. A sharp onset of highly direction emission from the cavity occurs at $j_p = 44$ mJ cm$^{-2}$, which corresponds to the lasing threshold. Inset: the spectral profile of emitted light is characterized by a narrow linewidth ($\Gamma$) of 2 meV. **c**, A horizontal intensity profile and a corresponding two-dimensional intensity profile of the output beam at a distance of 2.5 m from the cavity. The beam radius measured at the $1/e^2$ intensity level is 1.8 mm. **d**, The beam radius measured as a function of distance from the cavity (circles). The red line is a fit to a hyperbola. Here, $\theta = 0.66$ mrad (=0.038°) is the divergence half-angle, and $w_0 = 0.28$ mm is the beam waist radius. **e**, Spectrally tunable lasing spectra obtained with two samples of the type-(I+II) QDs having CdSe core radii 2.6 nm (same sample as in Figs. 2 and 3) and 2.3 nm. The latter sample has the following structure: CdSe ($r = 2.3$ nm)/ZnSe ($l = 2.2$ nm)/Cd$_{0.89}$Zn$_{0.11}$S ($h = 1.7$ nm)/ZnS ($d = 0.3$ nm). Using the larger-core QDs, the lasing line can be continuously tuned from 1.96 to 2.03 eV (634 to 610 nm) by varying the grating's blaze angle. With a smaller-core sample, we can tune emission further to the blue and reach spectral energies from 2.07 to 2.09 eV (599 to 594 nm). Inset shows higher resolution measurements of the lasing spectrum for the device containing a beam expander in the cavity. The observed linewidth is 380 μeV (=1.2 Å).

Supplementary Information for:

# Liquid-State Semiconductor Lasers Based on Type-(I+II) Colloidal Quantum Dots

Donghyo Hahm[1], Valerio Pinchetti[1], Clément Livache[1], Namyoung Ahn[1], Jungchul Noh[1], Xueyang Li[2], Jun Du[1,2], Kaifeng Wu[1,2], and Victor I. Klimov[1]*

[1]Nanotechnology and Advanced Spectroscopy Team, C-PCS, Chemistry Division, Los Alamos National Laboratory, Los Alamos, NM 87545, USA

[2]State Key Laboratory of Molecular Reaction Dynamics, Dalian Institute of Chemical Physics, Chinese Academy of Sciences, Dalian, Liaoning 116023, China

*klimov@lanl.gov

# Supplementary Note 1. Effective-mass calculations of electronic states

We utilized quantum-mechanical effective-mass calculations conducted using a custom Python code to obtain transition energies and wavefunctions of single excitons and biexcitons in type-(I+II) quantum dots (QDs). These calculations employed a finite difference mesh approach implemented using cylindrical coordinates. At each discretized mesh point, Schrödinger and Poisson equations were solved within the framework of the self-consistent field and effective mass approximation[1]. The calculations were carried out for the type-(I+II) QDs with the following structure: CdSe ($r$ = 2.6 nm)/ ZnSe ($l$ = 1.7 nm)/ CdS ($h$ = 2.2 nm)/ ZnS ($d$ = 0.3 nm) QDs. These dimensions corresponded to those of the QDs used in the majority of the experiments reported in the main article.

### A. Single excitons

A single-exciton wavefunction ($\Psi_X(r_e, r_h)$) was obtained from:

$$\Psi_X(r_e, r_h) = \psi_e(r_e)\psi_h(r_h) \quad (S1)$$

$$\left[-\frac{\hbar^2}{2m_e^*(r)}\nabla^2 + V_{CB}(r)\right]\psi_e(r) = E_e\psi_e(r) \quad (S2)$$

$$\left[-\frac{\hbar^2}{2m_h^*(r)}\nabla^2 + V_{VB}(r)\right]\psi_h(r) = E_h\psi_h(r) \quad (S3)$$

In the above equations, $\psi_e(r_e)$ and $\psi_h(r_h)$ are the electron and hole wave functions, respectively, $V_{CB}(r)$ and $V_{VB}(r)$ are the conduction- and valence-band confinement potentials, respectively, $m_e^*$ and $m_h^*$ are the electron and hole effective masses, respectively, and $E_e$ and $E_h$ are the electron and hole energy eigenvalues, respectively.

We permitted the electron wavefunctions ($\psi_e$) to adapt in response to the Coulombic attraction exerted by the hole, and vice versa. We employed the calculated electron and hole wavefunctions to obtain the Coulomb potentials ($\phi_{e,h}$) using the Poisson equation:

$$\nabla \cdot (\epsilon\epsilon_0 \nabla \phi_{e,h}(r)) = -q|\psi_{e,h}(r)|^2 \tag{S4}$$

where $\epsilon$ represents the permittivity of the material, $\epsilon_0$ is the vacuum permittivity, and $q$ is the elementary charge. Having obtained these potentials, we utilize them in the Schrödinger-Poisson equations for both the electron and the hole:

$$\left[-\frac{\hbar^2}{2m_e^*(r)}\nabla^2 + V_{CB}(r) + q\phi_h(r)\right]\psi_e(r) = E_e\psi_e(r) \tag{S5}$$

$$\left[-\frac{\hbar^2}{2m_h^*(r)}\nabla^2 + V_{VB}(r) - q\phi_e(r)\right]\psi_h(r) = E_h\psi_h(r) \tag{S6}$$

The solutions of eqs S5 and S6 were used in eq S4 to compute the 'corrected' Coulomb potentials, which then were again utilized in eqs S5 and S6 to obtain the next iteration for the wave functions. This iterative process continued until the energies of the electron and hole states converged. Typically, around four iterations were required to achieve convergence error $\Delta E_{e,h}/E_{e,h} \leq 0.1\%$. The exciton energy ($E_X$) was computed from:

$$E_X = E_e + E_h \tag{S7}$$

The conducted calculations resulted in the identification of two distinct near-band-edge exciton states for the type-(I+II) QDs with structure CdSe ($r$ = 2.6 nm)/ ZnSe ($l$ = 1.7 nm)/ CdS ($h$ = 2.2 nm)/ ZnS ($d$ = 0.3 nm). These states were classified as direct ($X_d$) and indirect ($X_i$) excitons. Their energies were: $E_d$ = 2.037 eV and $E_i$ = 1.988 eV. The corresponding wave functions are displayed in Supplementary Fig. 5.

## B. Biexcitons

The energies of three biexciton states ($X_iX_d$, $X_iX_i$, $X_dX_d$) were obtained using the results calculated for single excitons. To account for multi-carrier Coulomb interactions, we incorporated an additional interaction term:

$$E_{e,h,m-e,h,n} = \int \psi_{e,h,m}(r) q \phi_{e,h,n}(r) \psi_{e,h,m}(r)\, dV \tag{S8}$$

where *m* and *n* are the indices referring to a specific single-exciton state (denoted by i or d), from which a given electron or hole state originates.

The biexciton energies were calculated as follows:

$$E_{id} = E_i + E_d + E_{e,d-h,i} + E_{e,d-e,i} + E_{h,d-h,i} + E_{h,d-e,i} \tag{S9}$$

or

$$E_{id} = E_i + E_d + E_{e,i-h,d} + E_{e,i-e,d} + E_{h,i-h,d} + E_{h,i-e,d} \tag{S10}$$

$$E_{ii} = 2E_i + E_{e,i-h,i} + E_{e,i-e,i} + E_{h,i-h,i} + E_{h,i-e,i} \tag{S11}$$

$$E_{dd} = 2E_d + E_{e,d-h,d} + E_{e,d-e,d} + E_{h,d-h,d} + E_{h,d-e,d} \tag{S12}$$

These equations yield $E_{id}$= 4.024 eV, $E_{ii}$= 4.020 eV, and $E_{dd}$= 4.074 eV. Based on the calculations, the $X_iX_i$ and $X_iX_d$ biexcitons are energetically favored over the $X_dX_d$ biexciton. Therefore, in our analysis of experimental data, we account only for the lower-energy $X_iX_i$ and $X_iX_d$ states. The emission energies of these states can be found from:

$$h\nu_{id} = E_{id} - E_i \tag{S13}$$

$$h\nu_{ii} = E_{ii} - E_i \tag{S14}$$

These expressions yield $h\nu_{id} = 2.036$ eV and $h\nu_{ii} = 2.032$ eV. Both these spectral energies fall near the energy of the $X_d$ state (2.037 eV).

## Supplementary Note 2. Modeling of photoluminescence dynamics

Utilizing the three-level scheme depicted in the inset of Fig. 2c (main article), we formulated the following rate equations governing the temporal evolution of populations of the direct ($n_d$) and indirect ($n_i$) exciton states ($n_d + n_i = 1$):

$$\frac{dn_d}{dt} = -\frac{n_d}{\tau_d^*} + \frac{n_i}{\tau_{i-d}} \tag{S15}$$

$$\frac{dn_i}{dt} = -\frac{n_i}{\tau_i^*} + \frac{n_d}{\tau_{d-i}} \tag{S16}$$

where

$$\frac{1}{\tau_d^*} = \frac{1}{\tau_d} + \frac{1}{\tau_{d-i}} \tag{S17}$$

$$\frac{1}{\tau_i^*} = \frac{1}{\tau_i} + \frac{1}{\tau_{i-d}} \tag{S18}$$

The solutions of these coupled differential equations were presented as follows:

$$n_d = C_1 e^{-\frac{t}{\tau_{f,X}}} + C_2 e^{-\frac{t}{\tau_{s,X}}} \tag{S19}$$

$$n_i = C_3 e^{-\frac{t}{\tau_{f,X}}} + C_4 e^{-\frac{t}{\tau_{s,X}}} \tag{S20}$$

$$\frac{1}{\tau_{s(f),X}} = \frac{1}{2}\frac{1}{\tau_d^*}\left[(1+\beta) \pm \sqrt{(1+\beta)^2 - 4\beta(1-\alpha_1\alpha_2)}\right] \tag{S21}$$

where

$$\beta = \frac{\tau_d^*}{\tau_i^*} \tag{S22}$$

$$\alpha_1 = \frac{\tau_d^*}{\tau_{d-i}} \tag{S23}$$

$$\alpha_2 = \frac{\tau_i^*}{\tau_{i-d}} \tag{S24}$$

The dynamics of the intensity ($I$) of the combined photoluminescence (PL) emitted by the $X_i$ and $X_d$ states were computed from:

$$I(t) = \frac{n_d}{\tau_d} + \frac{n_i}{\tau_i} = \left(\frac{C_1}{\tau_d} + \frac{C_3}{\tau_i}\right)e^{-\frac{t}{\tau_{f,X}}} + \left(\frac{C_2}{\tau_d} + \frac{C_4}{\tau_i}\right)e^{-\frac{t}{\tau_{s,X}}} \tag{S25}$$

By considering the relationship between eigenvalues and eigenvectors of eqs S19 and S20, we obtained:

$$\frac{C_3}{C_1} = \tau_{i-d}\left(\frac{1}{\tau_d^*} - \frac{1}{\tau_{f,X}}\right) \tag{S26}$$

$$\frac{C_4}{C_2} = \tau_{i-d}\left(\frac{1}{\tau_d^*} - \frac{1}{\tau_{s,X}}\right) \tag{S27}$$

We further expressed the ratio of the amplitudes of the fast ($a_f$) and slow ($a_s$) PL components as

$$\frac{a_f}{a_s} = \frac{\frac{C_1}{\tau_d} + \frac{C_3}{\tau_i}}{\frac{C_2}{\tau_d} + \frac{C_4}{\tau_i}}, \tag{S28}$$

and the ratio of the time-integrated PL intensities originating from the $X_d$ ($A_d$) and $X_i$ ($A_i$) states as

$$\frac{A_d}{A_i} = \frac{\frac{C_1\tau_{f,X}}{\tau_d} + \frac{C_2\tau_{s,X}}{\tau_d}}{\frac{C_2\tau_{f,X}}{\tau_i} + \frac{C_4\tau_{s,X}}{\tau_i}} \tag{S29}$$

Then, we used eqs S21 and S25-S29 to model the measured PL dynamics. In this modeling, we used experimentally determined time constants of the fast ($\tau_{f,X}$ = 29 ns) and slow ($\tau_{s,X}$ = 697 ns) PL components, the measured ratio of their amplitudes ($a_f/a_s$ = 1.25) and the ratio of $A_d$ and $A_i$ (= 0.114) determined from the deconvolution of the experimental PL spectra (Supplementary Fig. 3).

Based on the best correspondence between the measured and computed dynamics, we obtained the following set of relaxation parameters of the excitonic system described by the scheme in the inset of Fig. 2c (main article): $\tau_d$ = 44 ns, $\tau_i$ = 735 ns, $\tau_{d-i}$ = 85 ns, and $\tau_{i-d}$ = 8800 ns.

## **Supplementary Note 3. Evaluation of critical QD concentrations**

The critical concentration of the QDs needed for laser action ($n_{QD,cr}$) was calculated from $n_{QD,cr} = (\gamma \tau_g)^{-1}$, where $\gamma = c\sigma_g/(\beta n_r)$ (ref. [2-4]). In these expressions, $\tau_g$ is the optical gain lifetime, $\sigma_g$ is the optical gain cross-section, $c$ is the speed of light, $n_r$ is the refractive index of the active medium, and $\beta$ is a dimensionless constant dependent on the specifics of the QD-optical-cavity system. Parameter $\beta$ is a multiplier that defines the relationship between a nominal stimulated emission build-up time ($\tau_{SE} = n_r/(cG_o)$; here $G_o$ is the saturated gain coefficient) and an actual build-up time of a cavity mode for a given laser configuration ($\tau_0$): $\tau_0 = \beta \tau_{SE}$. Previous modeling of microcavities utilized $\beta = 20$ (ref. 3). For a fairly long cavity (a few centimeters) used in the present study, we expect a slower build-up of a lasing mode which should translate in a larger value of $\beta$. In our estimations, we assumed that $\beta$ was 10-fold of that used in ref. 3, that is, $\beta = 200$. As shown below, this value yields a reasonable estimate of the lowest QD concentration required for liquid-state lasing. Indeed, using $\beta = 200$ along with $n_r = 1.496$ (for toluene), and $\sigma_g = 4.0 \times 10^{-16}$ cm$^2$ and $\tau_g = 2.4$ ns (inferred from transient absorption measurements), we obtain $n_{QD,cr}$ ~$10^{16}$ cm$^{-3}$ or ~18 µmol·L$^{-1}$. This estimated value of $n_{QD,cr}$ is close to the lowest QD concentration of 14 µmol·L$^{-1}$ for which we observed lasing with our type-(I+II) QD solution samples.

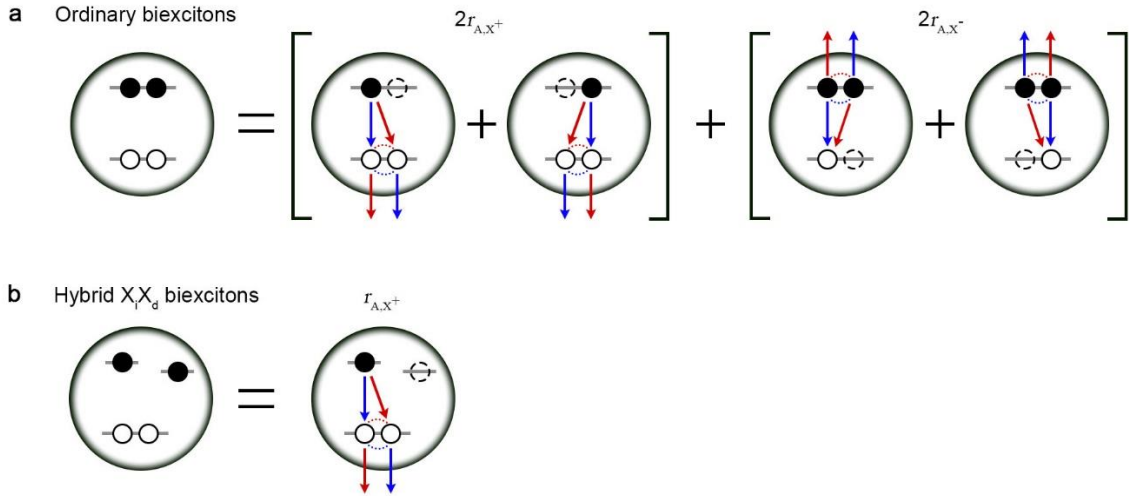

**Supplementary Fig. 1 | Application of the superposition principle to Auger decay of an ordinary (XX) biexciton and a hybrid ($X_iX_d$) biexciton. a**, According to the superposition principle, an Auger decay rate of any multicarrier state can be presented as a sum of rates of independent negative- and positive-trion Auger pathways ($r_{A,X^-}$ and $r_{A,X^+}$, respectively)[5-7]. In the case of an ordinary biexciton, there are 2 negative-trion and 2 positive-trion pathways. Hence, the total Auger rate is $r_{A,XX} = 2r_{A,X^-} + 2r_{A,X^+}$. **b**, In the case of a hybrid $X_iX_d$ biexciton, pathways involving an indirect transition can be neglected, which leaves operative only one negative-trion pathway. This yields $r_{A,XiXd} = r_{A,X^-}$.

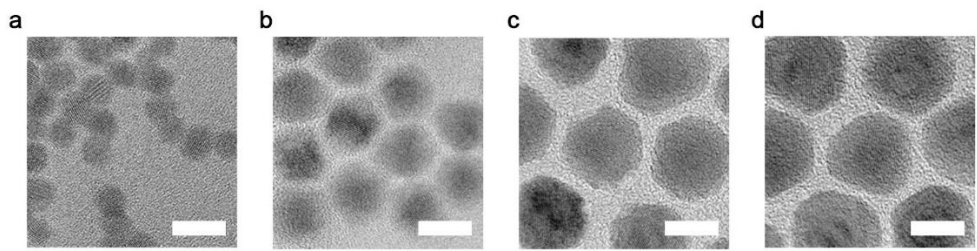

**Supplementary Fig. 2 | Transmission electron microscopy (TEM) images of type-(I+II) QDs and intermediate structures. a-d**, TEM images illustrating the progression of the synthesis of the type-(I+II) CdSe ($r$ = 2.6 nm)/ZnSe ($l$ = 1.7 nm)/CdS ($h$ = 2.2 nm)/ZnS ($d$ = 0.3 nm) QDs: (a) CdSe cores, (b) CdSe/ZnSe core/barrier structures, (c) CdSe/ZnSe/CdS core/barrier/interlayer structures, and (d) final CdSe/ZnSe/CdS/ZnS core/barrier/interlayer/shell QDs. The scale bars in the images correspond to 10 nm.

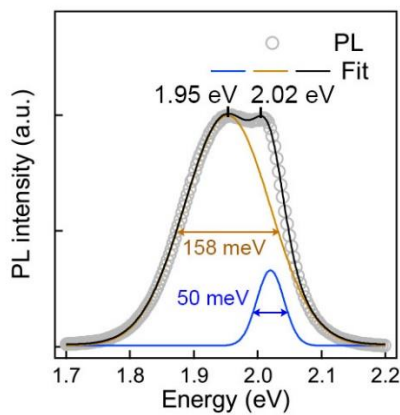

**Supplementary Fig. 3 | Double-band deconvolution of the PL spectrum of type-(I+II) QDs.** The PL spectrum of the type-(I+II) CdSe ($r$ = 2.6 nm)/ZnSe ($l$ = 1.7 nm)/CdS ($h$ = 2.2 nm)/ZnS ($d$ = 0.3 nm) QDs recorded using low-intensity (sub-single exciton) excitation (open grey circles) and the fit (black line) obtained via summation of the two Gaussian bands (gold and blue lines) whose peak positions and full widths at half-maximum (FWHM) are indicated in the figure.

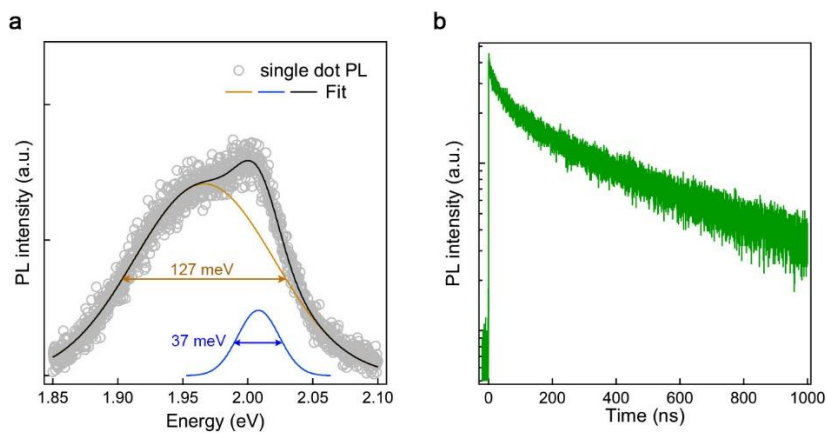

**Supplementary Fig. 4 | PL measurements of an individual type-(I+II) QD. a,** The measured PL spectrum (open grey circles) of a single type-(I+II) CdSe ($r$ = 2.6 nm)/ZnSe ($l$ = 1.7 nm)/CdS ($h$ = 2.2 nm)/ZnS ($d$ = 0.3 nm) QD fitted to a sum (black line) of the two Gaussian bands (cyan and orange lines). **b,** The corresponding PL dynamics recorded without spectral filtering. All data were collected using low-intensity sub-single-exciton excitation.

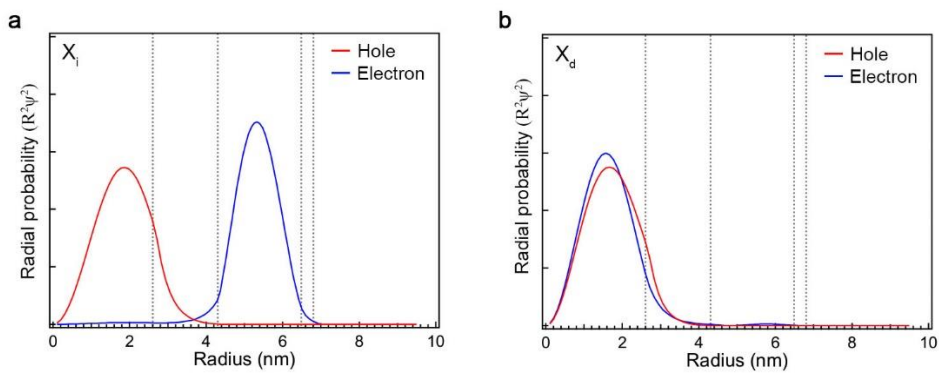

**Supplementary Fig. 5 | Radial electron and hole probabilities for the band-edge and the first excited quantized states. a,b**, The calculated electron (blue lines) and hole (red lines) radial probabilities for the lowest energy (**a**) and the first-excited (**b**) states of the type-(I+II) CdSe ($r$ = 2.6 nm)/ZnSe ($l$ = 1.7 nm)/CdS ($h$ = 2.2 nm)/ZnS ($d$ = 0.3 nm) QD. These two states correspond to the indirect ($X_i$) and direct ($X_d$) excitons, respectively.

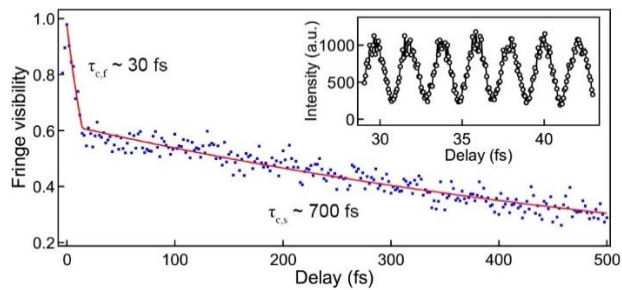

**Supplementary Fig. 6 | Temporal coherence measurements of a laser output using Michelson interferometry.** Fringe visibility as a function of time delay between the two arms of the interferometer (blue points) is fit to double-exponential decay (red line) with time constants 30 fs and 700 fs. The faster component arises from parasitic amplified spontaneous emission (ASE), while the slower component is due to lasing. The inset displays a representative interferogram.

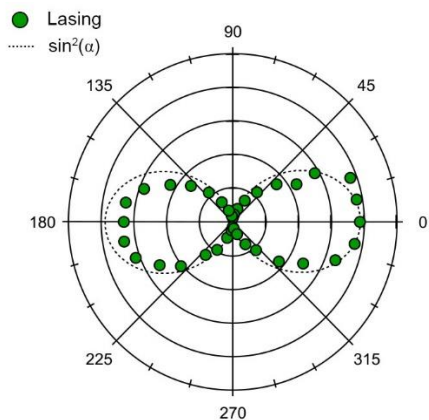

**Supplementary Fig. 7 | Polarization measurements of a laser output.** The emission intensity collected through a linear polarizer as a function of polarizer angle (green circles; shown using polar coordinates). The dotted line represents the ideal $\sin^2(\alpha)$ dependence expected for a perfectly linearly polarized output. These measurements indicate a high degree of linear polarization of 0.95.

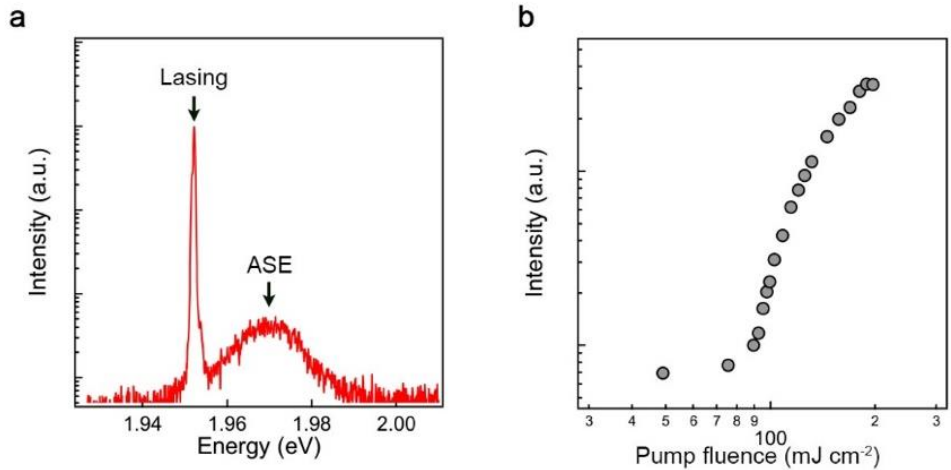

**Supplementary Fig. 8 | Lasing versus amplified spontaneous emission (ASE). a**, A representative spectrum of the liquid-state lasing device based on type-(I+II) CdSe ($r$ = 2.6 nm)/ZnSe ($l$ = 1.7 nm)/CdS ($h$ = 2.2 nm)/ZnS ($d$ = 0.3 nm) QDs. The narrow lasing line is at 1.952 eV and the broader ASE feature is at 1.97 eV. The amplitude of the ASE band is more than two orders of magnitude smaller than that of the lasing line (the 'y-axis' is logarithmic). The QD gain medium is pumped by 2.33-eV, 5-ns, 10-Hz pulses of the second harmonic of a Nd:YAG laser. **b,** The intensity of the ASE signal as a function of pump level indicates the ASE threshold of 89 mJ cm$^{-2}$. This is approximately twice as high as the lasing threshold.

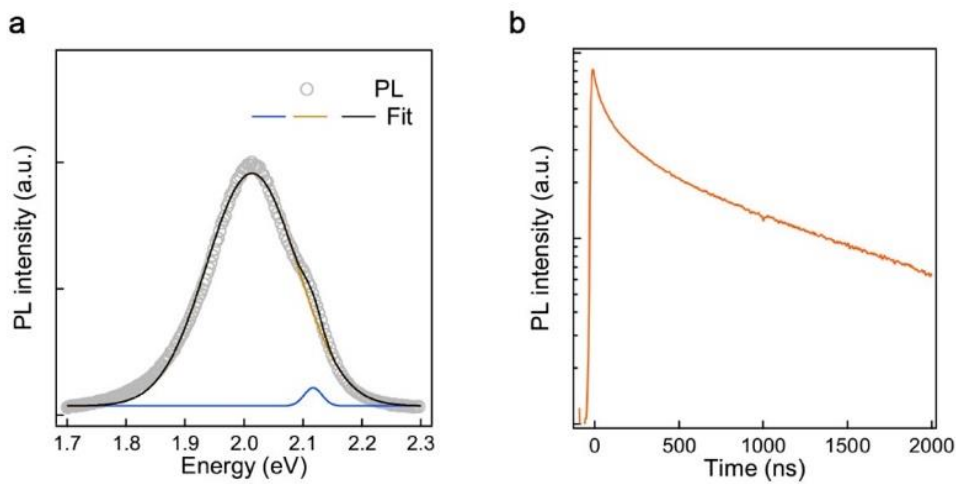

**Supplementary Fig. 9 | PL characteristics of CdSe/ZnSe/Cd$_{0.89}$Zn$_{0.11}$S/ ZnS QDs. a**, The PL spectrum of type-(I+II) CdSe ($r$ = 2.3 nm)/ ZnSe ($l$ = 2.2 nm)/ Cd$_{0.89}$Zn$_{0.11}$S ($h$ = 1.7 nm)/ ZnS ($d$ = 0.3 nm) QDs. The experimental data are shown by the open grey circles, while the double Gaussian fit is shown by the black line. Two individual Gaussian bands used in the fit are shown by the blue and orange lines. **b**, PL dynamics of the same sample collected using low-intensity excitation (the average per-dot exciton number is < 0.1). The sample was excited by 110-fs, 3.6-eV pulses.

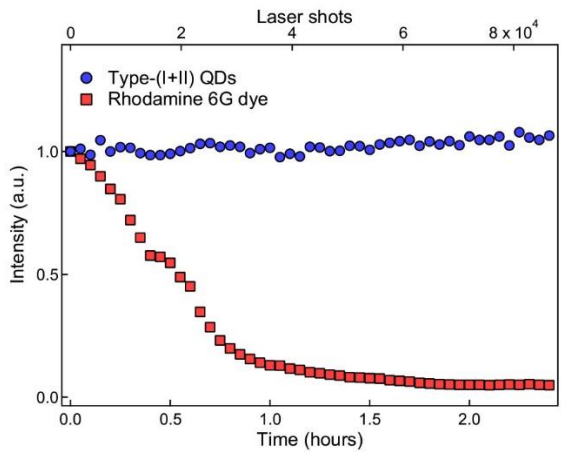

**Supplementary Fig. 10 | Comparison of operational stability of lasers based on type-(I+II) QDs and rhodamine 6G dye.** The intensity of a laser output for the gain medium based on the solution of the type-(I+II) CdSe ($r$ = 2.6 nm)/ZnSe ($l$ = 1.7 nm)/CdS ($h$ = 2.2 nm)/ZnS ($d$ = 0.3 nm) QDs (blue circles) versus rhodamine 6G (red squares). In both cases, the gain medium was excited by 532-nm, 5-ns, 10-Hz pulses of the second harmonic of the Nd:YAG laser. The QDs were dispersed in toluene, while rhodamine 6G was dissolved in a mixture of methanol and ethylene glycol (5:95 %, v/v) commonly used in commercial dye lasers. Both solutions were in a 'static form', that is, were not either stirred or flowed (or agitated in any other way) during the measurements.

1. Park, K., Deutsch, Z., Li, J. J., Oron, D. & Weiss, S. Single molecule quantum-confined Stark effect measurements of semiconductor nanoparticles at room temperature. *ACS Nano* **6**, 10013-10023 (2012).
2. Klimov, V. I. *et al.* Optical gain and stimulated emission in nanocrystal quantum dots. *Science* **290**, 314-317 (2000).
3. Park, Y.-S., Bae, W. K., Baker, T., Lim, J. & Klimov, V. I. Effect of Auger recombination on lasing in heterostructured quantum dots with engineered core/shell interfaces. *Nano Lett.* **15**, 7319-7328 (2015).
4. Park, Y.-S., Roh, J., Diroll, B. T., Schaller, R. D. & Klimov, V. I. Colloidal quantum dot lasers. *Nat. Rev. Mater.* **6**, 382-401 (2021).
5. Park, Y.-S., Bae, W. K., Padilha, L. A., Pietryga, J. M. & Klimov, V. I. Effect of the core/shell interface on Auger recombination evaluated by single-quantum-dot spectroscopy. *Nano Lett.* **14**, 396-402 (2014).
6. Park, Y.-S., Bae, W. K., Pietryga, J. M. & Klimov, V. I. Auger recombination of biexcitons and negative and positive trions in individual quantum dots. *ACS Nano* **8**, 7288-7296 (2014).
7. Wu, K., Lim, J. & Klimov, V. I. Superposition principle in Auger recombination of charged and neutral multicarrier states in semiconductor quantum dots. *ACS Nano* **11**, 8437-8447 (2017).